\documentclass[nofootinbib,twocolumn,superscriptaddress,pre,aps,reprint,showpacs,showkeys]{revtex4-1}
\usepackage{bm,amsmath,color}
\usepackage{graphicx}

\usepackage{bm}
\usepackage{amsmath,color}
\usepackage{amssymb}
\usepackage{lipsum}
\usepackage{ulem}
\usepackage{setspace}

\def\be{\begin{equation}}
\def\en{\end{equation}}
\def\p{\partial}
\def\d{\mathrm d}

\begin{document}
\title{Dynamics of a Bilayer Membrane Coupled to a Two-dimensional Cytoskeleton: Scale Transfers of Membrane Deformations}

\author{Ryuichi Okamoto}
\affiliation{Department of Chemistry, Graduate School of Science and Engineering, Tokyo Metropolitan University, Tokyo 192-0397, Japan}

\author{Shigeyuki Komura}
\affiliation{Department of Chemistry, Graduate School of Science and Engineering, Tokyo Metropolitan University, Tokyo 192-0397, Japan}

\author{Jean-Baptiste Fournier}
\affiliation{Universit\'e Paris Diderot, Sorbonne Paris Cit\`e, Laboratoire Mati\`ere et Syst\`emes Complexes (MSC), UMR 7057 CNRS, F-75205 Paris, France}

\date{\today}
\begin{abstract}
We theoretically investigate the dynamics of a floating lipid bilayer membrane coupled with a two-dimensional cytoskeleton network, taking into explicitly account the intermonolayer friction, the discrete lattice structure of the cytoskeleton, and its prestress. The lattice structure breaks lateral continuous translational symmetry and couples Fourier modes with different wavevectors. It is shown that within a short time interval a long-wavelength deformation excites a collection of modes with wavelengths shorter than the lattice spacing. These modes relax slowly with a common renormalized rate originating from the long-wavelength mode. As a result, and because of the prestress, the slowest relaxation is governed by the intermonolayer friction. Reversely, and most interestingly, forces applied at the scale of the cytoskeleton for a sufficiently long time can cooperatively excite large-scale modes.

\end{abstract}
\pacs{87.16.D-, 87.16.Ln, 68.03.Cd}
\maketitle

\section{Introduction}
In biological materials, much attention has been paid to the dynamics from the viewpoint of nonequilibrium physics, because of the high complexity of composition, hydrodynamic interactions, active components, etc.~\cite{Mizuno,Weitz,Mikhailov}.
In particular, shape relaxation and fluctuations of lipid bilayer membranes have intensively been studied both in, or near, equilibrium \cite{Brochard, Sackmann, Seifert} and far from equilibrium~\cite{Prost, Garcia,Turlier, BetzPNAS, ParkPNAS, YasudaPRE}. 
The dynamics of a bilayer membrane is determined by many factors, such as the viscosity of the surrounding fluid~\cite{Brochard}, the membrane bending rigidity, the inter-monolayer friction caused by relative lateral motions of two monolayers~\cite{Seifert, JBPRL2}, and possibly by active inclusions~\cite{Prost}.
Red blood cell (RBC) membranes have further complexity because of the cytoskeleton which is attached to the lipid bilayer. It has been argued that the cytoskeleton plays crucial roles both in the statics and the dynamics, e.g., drastic effective tension increase in equilibrium \cite{GovPRL,JBPRL,JBEPL,Popescu}, tension decrease in the presence of ATP \cite{BetzPNAS,  Turlier}, and enhanced non-equilibrium fluctuations on the scale of the cytoskeleton mesh size \cite{ParkPNAS}.

In RBCs, the cytoskeleton consists of spectrin filaments forming a pre-stressed~\cite{Turlier} two-dimensional (2D) triangular lattice with a protein at each vertex embedded in the bilayer membrane. The lattice spacing $a\approx 100\,\mathrm{nm}$ is quite large, and therefore what matters in the membrane collective dynamics is not only
the modes whose wavelengths are much larger than $a$, but also those having wavelengths smaller than $a$. To understand simultaneously the dynamics on such a wide range of spatial scales, we need to take into explicitly account the discrete nature of the lattice structure. The latter breaks lateral continuous translational symmetry, giving rise to a coupling between modes on different length scales and thus to a rich dynamical behavior.

This paper is organized as follows. In Sec.~\ref{FE}, our model free energy is constructed on the basis of the previous theories for a bilayer membrane (without a cytoskeleton) \cite{Seifert} and for a pre-stressed 2D cytoskeleton coupled to a membrane \cite{JBEPL}. In Sec.~\ref{DE}, following Ref.~\cite{Seifert}, the hydrodynamic equations are introduced, where we take into account the hydrodynamic flows of the surrounding fluid and of the monolayers, inter-monolayer friction between the monolayers. Then we obtain the coupled equations for the membrane variables by integrating out the flow velocity fields. In Sec.~\ref{result}, we discuss how the cytoskeleton alters the dynamics of the membrane, when initially a large-scale deformation is imposed, and when force(s) are applied to small-scale mode(s) for a long time ($\gtrsim 10$ ms). Section \ref{summary} is devoted for discussion and summary. We also present our detailed calculations in the Appendices.

\section{Free energy\label{FE}}
We consider out-of-plane deformations of a RBC membrane patch described by its height $h(x,y)$ above a reference plane $z=0$. Our model free energy is given by $F=F_0+F_{\rm c}$ as follows. We take into account the areal compression which is necessarily coupled with $h$ due to the finite thickness $d\approx 1\, {\rm nm}$ of the monolayers~\cite{Seifert}. Then the bilayer membrane free energy $F_0$ is given by
\begin{align}
F_0=\int \d ^2x\,\Big[ &\frac{\kappa}{2} (\nabla^2 h)^2+\frac{\sigma}{2}(\nabla h)^2 \nonumber \\&
+\frac{k}{2}\sum_{\epsilon=\pm}\,[\rho^\epsilon+\epsilon d(\nabla^2h)]^2 \Big], \label{F0}
\end{align}
where $\kappa$ is the bare bending rigidity, $\sigma$ the bare tension, $k$ the areal compression modulus and $\nabla=(\p_x,\p_y)$. In the above, $\rho^+$ (resp.~$\rho^-$) denotes the dimensionless projected excess lipid density in the upper (resp.~lower) monolayer \cite{Seifert}. 

Note that in our paper, the surface tension is not considered as a constant. The quantity $\sigma$ in Eq.~(\ref{F0}) is only the background tension for a flat membrane with homogeneous, reference lipid density ($\rho^\pm=0$). The actual tension fluctuates about the zeroth-order tension $\sigma$  according to a model that is closely related to the area-difference-elasticity model \cite{Miao}. Indeed, the term proportional to $k$ in Eq.~(\ref{F0}), that involves the variables $\rho^+$ and $\rho^-$ is the excess energy associated with a local compression or dilation of the lipids in each monolayer. The actual tension (without the cytoskeleton) is $ \sigma + k(\rho^+ + d \nabla^2 h)$ and $ \sigma + k(\rho^- - d \nabla^2 h)$ in the upper and lower monolayers, respectively.

\begin{figure}
\includegraphics[scale=0.65]{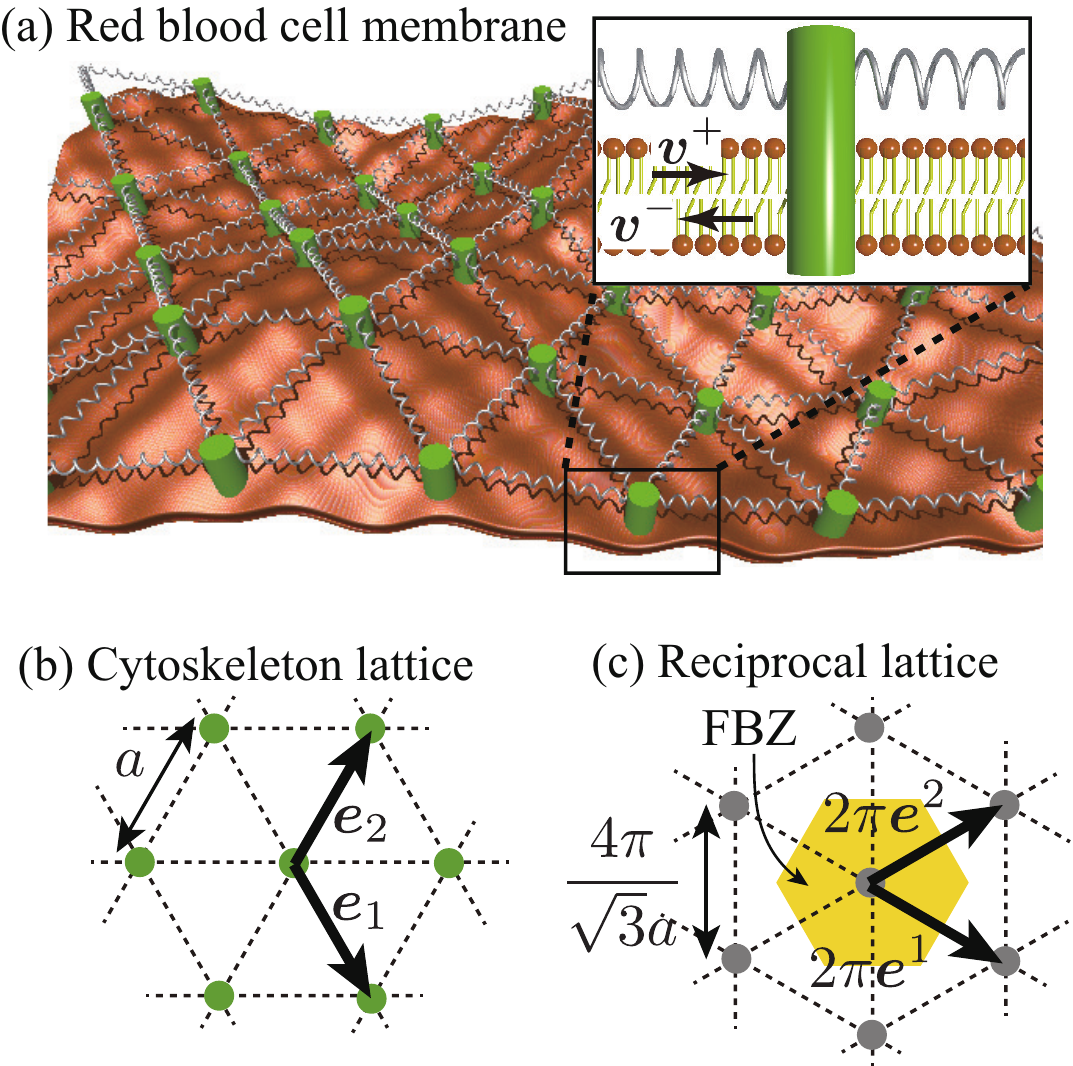}
\caption{(a) Illustration of the model red blood cell (RBC) membrane. (b) Cytoskeletal network. (c) Reciprocal lattice. The yellow region represents the first Brillouin zone (FBZ).}
\label{lattice}
\end{figure}

The other contribution, $F_{\rm c}$, arises from the membrane--cytoskeleton coupling. We assume, to simplify, that the cytoskeleton network is a uniform triangular lattice without defects. An anchoring protein at each lattice site is embedded in the membrane and interacts with its nearest neighbor through an effective spring of relaxed length $a_0$ and stiffness $k_{\rm s}$ (Fig.~\ref{lattice}a). In the ground state ($\rho=h=0$), the network forms a regular triangular lattice, and the lattice points $\{{\bm R}_\ell\}$ are expressed in terms of the primitive lattice vectors ${\bm e}_\alpha$ $(\alpha=1,2)$ as
${\bm R}_\ell=R_\ell^\alpha{\bm e}_\alpha$
with $R_\ell^\alpha\in\mathbb{Z}$ positive or negative integers (see Fig.~\ref{lattice}b). The lattice spacing is $a=|{\bm e}_\alpha|$. 
If the out-of-plane deformation of the membrane is sufficiently small, $F_{\rm c}$ is given by \cite{JBEPL}
\begin{align}
F_{\rm c}=\frac{\nu}{4} \sum_{\ell} \sum_{\bm n}\,[h({\bm R}_\ell)-h({\bm R}_\ell+{\bm n})]^2, \label{F_lattice}
\end{align}
where $\nu=k_{\rm s}(1-a_0/a)$ is the effective stiffness of the harmonic potentials associated with the out-of-plane deformations, and $\sum_{\bm n}$ denotes the sum over the nearest neighbor sites. Note that $\nu$ is nonzero only if the lattice is pre-stressed ($a\ne a_0$). In Ref.~\cite{Turlier}, as a result of fitting their experimental data, it has been shown that the cytoskeleton in healthy RBCs is naturally stretched (by about $4\%$) while the bare membrane tension can be negative.
Let us introduce the in-plane Fourier transform  as ${\cal F}_{\bm q}[\cdots] \equiv\int {\rm d}^2x\, (\cdots)e^{-i{\bm q}\cdot{\bm x}}$ and the reciprocal vectors ${{\bm e}^\alpha}$ satisfying ${\bm e}^\alpha\cdot{\bm e}_\beta=\delta^\alpha_\beta$, the Kronecker delta (see Fig.~\ref{lattice}c). Because $F_{\rm c}$ breaks the {\it lateral continuous translational symmetry}, modes with different wavevectors are coupled to one another. As shown in Appendix \ref{couple_mode}, the modes coupled to a given $\bm q$ belong to the subset 
\begin{align}
Q_{\bm q}=\{ {\bm q}+2\pi m_\alpha {\bm e}^\alpha | m_\alpha  \in \mathbb{Z} \}.
\end{align} 

\section{Dynamic equations \label{DE}}
Following Seifert and Langer~\cite{Seifert}, we regard each monolayer as a compressible 2D fluid having the shear viscosity $\mu$ and the bulk viscosity $\zeta$. The upper and lower monolayers can have different fluid velocities, ${\bm v}^+$ and ${\bm v}^-$, respectively (Fig~\ref{lattice}a). The full dynamic equations consist of (i) lateral force balance for each monolayer, (ii) force balance normal to the bilayer, and (iii) the continuity equation for lipids in each monolayer.  

We use the Stokes equation for the solvent velocity field ${\bm V}$ and the pressure field $p$, with the shear viscosity  $\eta$:
 \begin{align}
\eta \hat\nabla ^2 {\bm V}-\hat\nabla p=0, \quad \hat\nabla\cdot {\bm V}=0, \label{Stokes}
 \end{align}
 where $\hat\nabla=(\p_x,\p_y,\p_z)$ is the 3D nabla operator.
No-slip boundary condition is employed at the membrane surface, 
$v^\pm_i=V_i \ (i=x,y)$ and $V_z=\p h/\p t$ at $z\to 0^\pm$. We also impose ${\bm V}\to 0$ and $p\to p_0$ as $z\to \pm \infty$.
The 2D viscous stress tensors in the monolayers are given by
\begin{align}
\tau_{ij}^\pm=\mu(\p_i v_j^\pm+\p_j v_i^\pm )+(\zeta-\mu)\delta_{ij}\nabla \cdot {\bm v}^\pm, 
\end{align}
where the superscript ``$+$" (resp.~``$-$") denote the upper (resp.~lower) monolayer. 
Then the lateral force balance equation in each monolayer reads 
\begin{align}
-\p_i \left(\frac{\delta F}{\delta\rho^\pm}\right)+\p_j\tau_{ij}^\pm\pm T_{iz}^\pm\mp b(v^+_i-v^-_i)=0, \label{lateral}
\end{align}
where $T_{ij}^+$ (resp. $T_{ij}^-$) is the stress tensor $T_{ij}=-p\delta_{ij}+\eta (\p_iV_j+\p_jV_i)$ in the solvent fluid evaluated at $z\to0^+$ (resp. $z\to 0^-$). The last term is due to the inter-monolayer friction, with  the friction coefficient $b$ \cite{Seifert}. In the normal direction, the force exerted by the surrounding fluid is balanced with the restoring force of the membrane,
\begin{align}
T^+_{zz}-T^-_{zz}=\frac{\delta F}{\delta h}. \label{vertical}
\end{align}
 At linear order in ${\bm v}^\pm$ and $\rho^\pm$, which are both considered to be small, the continuity equation in each monolayer is given by 
 \begin{align}
 \frac{\p \rho^\pm}{\p t}\simeq -\nabla\cdot {\bm v}^\pm. \label{continuity}
 \end{align}

The velocities, ${\bm V}$ and ${\bm v}^\pm$, can be eliminated from the dynamic equations by integrating the Stokes equations along $z$ for each mode $\bm q$ (see Appendix \ref{elimination_velo}). This yields coupled linear equations for $\hat{h}\equiv h/d$ and $\rho\equiv (\rho^+-\rho^-)/2$:
\begin{align}
&4\eta d^2 q \frac{\p \hat{h}({\bm q},t)}{\p t}=-{\cal F}_{\bm q}\left[ \frac{\delta F}{\delta\hat{h}}\right] +u_h({\bm q},t), \label{EQ1}\\
&\frac{2c(q)}{q^2} \frac{\p \rho({\bm q},t)}{\p t}=-{\cal F}_{\bm q}\left[ \frac{\delta F}{\delta\rho}\right] +u_\rho({\bm q},t), \label{EQ2}
\end{align}
where $c(q) = 2b+2\eta q+(\mu+\zeta)q^2$ with $q=|\bm q|$. 
We have added $u_h$ and $u_\rho$, representing external forces applied mechanically (e.g., by active molecules) which act on the variables $\hat{h}$ and $\rho$, respectively. 
Since ${\cal F}_{\bm q} [\delta F /\delta h]$ includes $\hat{h}({\bm q}')$ for $\forall {\bm q}' \in Q_{\bm q}$, these equations actually consist of sets of coupled equations for the variables $\{\hat{h}({\bm q}'), \rho({\bm q}')\}$ in each set $Q_{\bm q}$ (see Appendix \ref{couple_mode}).

Without the cytoskeleton, the modes for different wavevectors are not coupled in Eqs.~(\ref{EQ1}) and (\ref{EQ2}). Then, $\hat{h}({\bm q})$ and $\rho({\bm q})$ exhibit two relaxation rates, $\gamma_+^{(0)}(q)>\gamma_-^{(0)}(q)$, associated with some linear combinations of $\hat{h}({\bm q})$ and $\rho({\bm q})$.
Seifert and Langer discussed these relaxation modes for vanishing tension~\cite{Seifert}. 
They found a crossover wavenumber $q_{\rm c}= 2\eta k/(b\tilde\kappa)\approx 4.4\times 10^6\, {\rm m}^{-1}$, at which the relaxation behavior of the membrane changes qualitatively. Here we set $\kappa=2\times 10^{-20}\,{\rm J} $ as in \cite{JBPRL} (the value of $\kappa$ measured in experiments lies in quite a wide range, $1$ to $30\times 10^{-20}\, {\rm J}$ \cite{Brochard,BetzPNAS,Turlier, Popescu}, but the following results remain almost unchanged even with these different values). For large scales satisfying $q\ll q_{\rm c}$,  the rates correspond to $\rho$ relaxing quickly followed by $h$ relaxing slowly with $\rho$ being slaved~\cite{JBNLM}. For small scales, $q\gg q_{\rm c}$, conversely, they correspond to $h$ relaxing quickly followed by $\rho$ relaxing slowly with $h$ being slaved. Hence, the dynamics on the small scales is dominated by the inter-monolayer friction, whereas that on the large scales is dominated by the solvent viscosity.
In the presence of tension, their results hold for $\sigma\ll\sigma_{\rm c}\equiv(2\eta k)^2/(\tilde\kappa b^2)$, except at very large scales (see Refs.~\cite{JBNLM,OkamotoEPJE, Sachin} and Appendix \ref{perturbation}).
However, for $\sigma\gtrsim\sigma_{\rm c}$ the dynamics is dominated at all scales by the inter-monolayer friction, with $\gamma_+^{(0)}\simeq(\sigma q+\kappa q^3)/(4\eta)>\gamma_-^{(0)}\approx kq^2/(2b)$~\cite{JBNLM}.

\section{results \label{result}}
\subsection{Relaxation of a large-scale deformation}
The cytoskeleton shifts the mode relaxation rates by an amount that depends on the prestress $\sim\!\nu$, and at the same times it couples all the modes belonging to a common set $Q_{\bm q}$. Let us first discuss how the rates of the large scale modes, with $q\ll q_{\rm c} \ll 2\pi /a$, are shifted by the cytoskeleton.
For such modes, the dependence on the direction of $\bm q$ is negligible. In the following, analytical expressions will be given systematically at first-order in a perturbative expansion in power series of $\nu$ (see Appendix \ref{perturbation}). The parameter values used in the following numerical calculations are summarized in Table \ref{TabPara}.
\begin{table*}[tbh]
\caption{
List of the parameter values used in numerical calculations.
\label{TabPara}}
\begin{ruledtabular}
\begin{tabular}[t]{c c c c c c c c c}
$\sigma$  \quad & $\kappa$\ \ \quad  & $k$\   \quad & $d$\  \quad & $\nu$ & $a$ & $\eta$ & $b$\  \quad  & $\mu +\zeta $ \quad\\
 $\mathrm{N/m}\quad$ & $\mathrm{J}$\quad & $\mathrm{N}/\mathrm{m}$ \quad & $\mathrm{m}$\quad & $\mathrm{N/m}$ & $\mathrm{m}$ & $\mathrm{J}\, \mathrm{s}/\mathrm{m}^3\quad $&$\mathrm{J}\, \mathrm{s}/\mathrm{m}^4\quad$&$\mathrm{J}\, \mathrm{s}/\mathrm{m}^2$\\
\hline
$10^{-11}$ & $2\times 10^{-20} $ & $7 \times 10^{-2}$ & $10^{-9} $ \ & $10^{-6}$ & $10^{-7}$ & $10^{-3}$&$2\times 10^{8}$&$2\times 10^{-9}$\\
\end{tabular}
\end{ruledtabular}
\end{table*}

From the dynamic equations (\ref{EQ1})--(\ref{EQ2}), we find that the rates of the large-scale modes, $\gamma_+>\gamma_-$, are shifted according to
\begin{align}
\gamma_+\simeq \frac{\sigma_{\rm eff}q+\kappa_{\rm eff}q^3}{4\eta}, \quad
\gamma_- \simeq \Big( \frac{k}{2b}+ \frac{ 3^{1/2} d^2 b}{4 \eta^2}\nu \Big)q^2,
\label{eigenvalues1}
\end{align}
where $\sigma_{\rm eff}=\sigma+3^{1/2}\nu $ and $\kappa_{\rm eff}=\kappa-3^{1/2}\nu a^2/16$ are the tension and the bending rigidity renormalized by the cytoskeleton~\cite{JBEPL,Turlier}.
Note that the fast and slow rates have been exchanged with respect to their bare value $\gamma_+^{(0)}\simeq kq^2/(2b)>\gamma_-^{(0)}\simeq (\sigma q+\kappa q^3)/(4\eta)$ because we anticipate $\sigma_\mathrm{eff}\gtrsim\sigma_{\rm c}>\sigma$ for large enough $\nu$. 
The other rates $\gamma_\pm(q')$ of the subset ${\bm q}'\in Q_{\bm q}$ associated with ${\bm q}$ are also shifted from their bare values $\gamma_\pm^{(0)}(q')$; they correspond to wavelengths comparable to or smaller than the cytoskeleton mesh size and are much faster than those in Eq.~(\ref{eigenvalues1}). Our detailed calculations show that the shifts of these rates are small (see Appendix \ref{perturbation}), but this does not mean that the small-scale modes are not affected by the cytoskeleton

\begin{figure}
\includegraphics[scale=0.62]{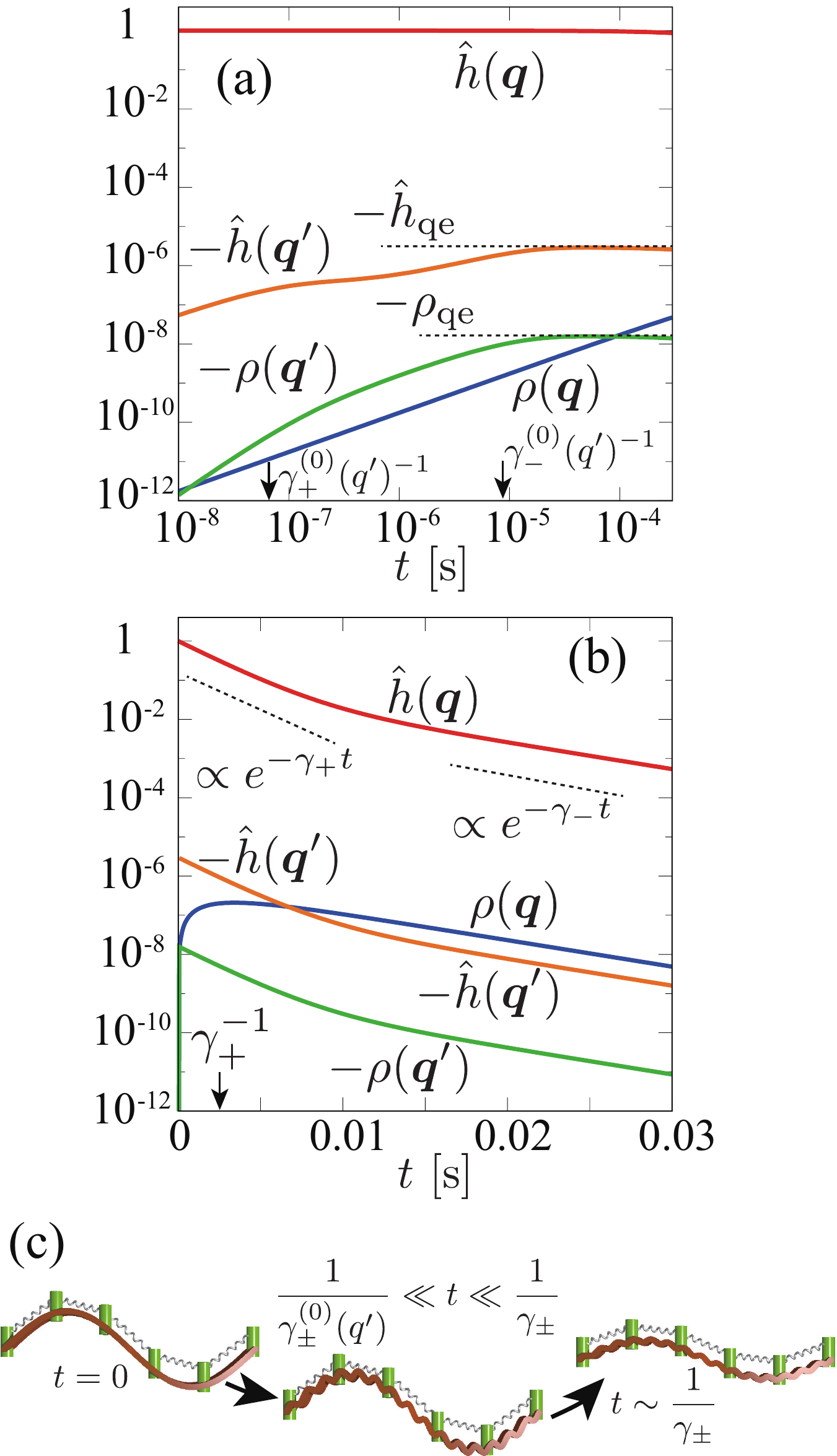}
\caption{Relaxation of a (normalized) large-scale deformation $\hat{h}({\bm x})=e^{i{\bm q}\cdot{\bm x}}$. (a) Short-time behavior of the large-scale modes $\hat h({\bm q})$ and $\rho({\bm q})$, and that of the small-scale modes $\hat h({\bm q}')$ and $\rho({\bm q}')$ with ${\bm q}'={\bm q}+2\pi{\bm e}^2$.  (b) Long-time behavior of the same variables. (c) Schematic pictures of the process. Although the amplitude of the excited small-scale mode is small, a large effective tension $\sigma_{\rm eff}$ changes qualitatively the large-scale dynamics such that it is dominated by the inter-monolayer friction. }
\label{evol1} 
\end{figure}

The relaxation of a large scale mode $\bm q$ excites all the small scale modes in $Q_{\bm q}$ (Fig.~\ref{evol1}). To investigate this effect, we set the initial condition $\hat h({\bm x})=e^{i{\bm q}\cdot {\bm x}}$ and  $\rho({\bm x})=0$ with $q=10^6\,\mathrm{m}^{-1}$ in the direction ${\bm q}/q=(\sqrt{3}/2,1/2)$, and we integrate numerically the dynamical equations up to the cutoff $20\pi/a$, of the order of the inverse membrane thickness.  Experiments indicate $\sigma_\mathrm{eff}\approx\nu\approx10^{-7}$--$10^{-5}\,\mathrm{N/m}$ with $\sigma$ very small or even negative~\cite{Turlier,Sackmann,Garcia, Popescu}. Accordingly, besides the values already given, we set $\sigma=10^{-11}\,\mathrm{N/m}$ and $\nu=10^{-6}\,\mathrm{N/m}$, yielding $\kappa_{\rm eff}=1.9\times 10^{-20}\,\mathrm{J}$ and $\sigma_{\rm eff}=1.73\times 10^{-6}\,\mathrm{N/m}$. 
We study the coupled evolution of $\hat h$ and $\rho$ for ${\bm q}$ and for small-scale modes ${\bm q}'\in Q_{\bm q}$. In Fig.~\ref{evol1}, we present as an example only ${\bm q}'={\bm q}+2\pi{\bm e}^2\simeq 2\pi{\bm e}^2$, as we find the other small-scale modes in $Q_{\bm q}$ also exhibit a similar behavior.
In the short time interval $0<t\ll1/\gamma_\pm(q')\approx1/\gamma_\pm^{(0)}(q')$, the small-scale modes $\hat h({\bm q}')$ and $\rho ({\bm q}')$, that are initially zero, are excited, while $\hat{h}({\bm q})$ almost remains unchanged (Fig.~\ref{evol1}a and c). All the excited small-scale modes rapidly approach their respective \textit{quasi-equilibrium states} $\hat{h}_{\rm qe}$ and $\rho_{\rm qe}$, which minimize the free energy for a fixed value of $\hat{h}({\bm q})$, given by
\begin{align}
\hat{h}_{\mathrm{qe}}({\bm q}';\hat{h}({\bm q}))=
\frac{\rho_{\mathrm{qe}}({\bm q}';\hat{h}({\bm q}))}{d^2q'^2} \simeq -\frac{2\nu\hat{h}({\bm q})K_{\bm q}}{\sqrt{3}\kappa a^2 q'^4}, \label{quasi_eq}
\end{align}
where $K_{\bm q}=\sum_{\bm n} (1-e^{i{\bm q}\cdot {\bm n}})$.  Figure \ref{evol1}b illustrates the long time evolution of the system. For $1/\gamma_\pm^{(0)}(q')\ll t \lesssim \gamma_+^{-1}$, $\hat h({\bm q})$, $\hat h({\bm q}')$ and $\rho({\bm q}')$ decay with the common rate $\gamma_+$. Then, around $t \approx \gamma_+^{-1}$, $\hat h({\bm q})$ follows the dynamical quasi-equilibrium value $\hat h_{\rm qe}^\rho ({\bm q};\rho({\bm q},t))$ that minimizes the free energy at fixed $\rho({\bm q},t)$. Finally, for $t \gg \gamma_+^{-1}$, all the modes decay with the common rate $\gamma_-$, with $\hat h({\bm q}')$, $\rho({\bm q}')$ and $\hat h({\bm q})$ following their respective dynamical quasi-equilibrium values $\hat h({\bm q}')\simeq \hat h_{\rm qe}({\bm q}';\hat h_{\rm qe}^\rho({\bm q}))$, $\rho({\bm q}')\simeq \rho_{\rm qe}({\bm q}';\hat h_{\rm qe}^\rho({\bm q}))$ and $\hat h({\bm q})\simeq \hat h_{\rm qe}^\rho({\bm q};\rho({\bm q},t))$ (Fig.~\ref{evol1}b and c). 
Suppose in Fig.~\ref{evol1} the initial amplitude of $h({\bm q})$ is comparable with the mode wavelength, $2\pi /q\approx 6.3\,\mu{\rm m}$. Then the amplitude of the excited small-scale mode $h({\bm q}')$ is about $2.4\times 10^{-5}\,\mu{\rm m}$, which is much smaller than the mode wavelength $2\pi /q'\approx 8.6 \times 10^{-2} \,\mu{\rm m}$, and may not be observable in experiments. This is because an energy cost to make a deformation with amplitude $q'^{-1}$ at the small-scale $q'$ is larger than to make a deformation with amplitude $q^{-1}$ at the large-scale $q$. Nevertheless, we notice that the cytoskeleton alters qualitatively the large-scale dynamics; because of the cytoskeleton that yields $\sigma_\mathrm{eff}\sim \sigma_{\rm c}\approx 3\times10^{-6}\,\mathrm{N/m}$, {\it the slowest relaxation process is dominated by the large-scale compression mode} $\rho({\bm q})$ limited by the inter-monolayer friction.

\subsection{Large-scale deformation induced by small-scale deformation via the cytoskeleton}
We have seen that large-scale deformations excite the modes whose scales are comparable or smaller than the cytoskeleton mesh. Now a question arises. Can small scale deformations excite large scale ones? If yes, is the amplitude of the excited modes large enough to be observable? The answer is no, if there are no applied forces. This is because the small-scale modes are much faster than the large-scale ones, so that the small-scale modes rapidly relax before large-scale modes are excited. However, if we \textit{keep} applying forces \textit{only} to the small-scale modes for a time longer than the relaxation time of the large-scale modes, the latter will be excited \textit{via} the cytoskeleton. Furthermore, if forces are applied to many small-scale modes, the amplitude of the excited large-scale mode can be noticeably large. In fresh RBCs, active molecules could be the source of these forces, as their characteristic time is of the order of $1\,\mathrm{s}$~\cite{BetzPNAS, Turlier} which is much larger than the typical relaxation time of the modes for $q\sim 10^6\,\mathrm{m}^{-1}$. It was further proposed that the active force is particularly enhanced on the scales of the cytoskeleton mesh \cite{ParkPNAS}.

\begin{figure}
\includegraphics[scale=0.53]{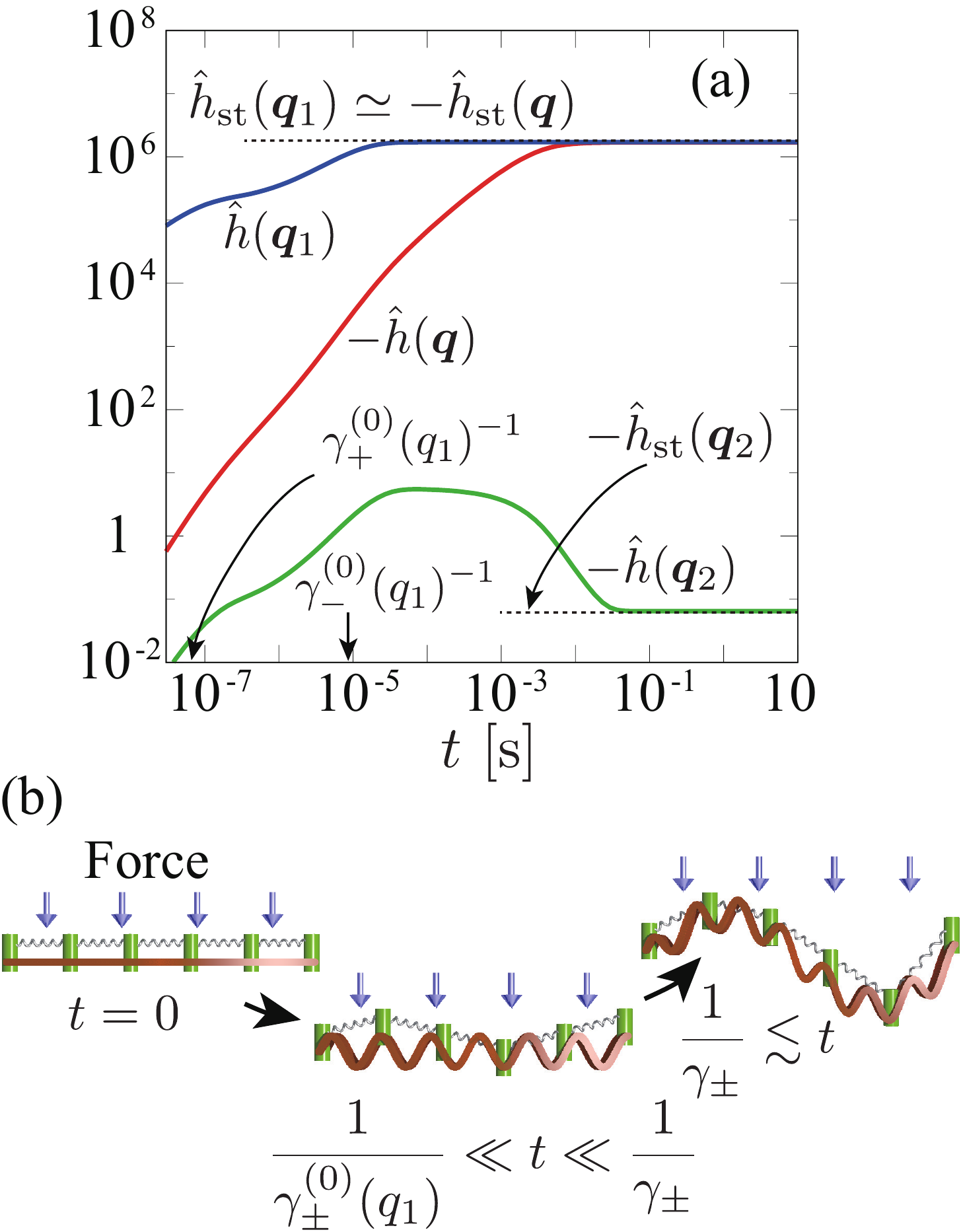}
\caption{Time evolution of the membrane shape under a constant force $\bar u$ applied to the small-scale mode $\hat{h}({\bm q}_1)$, with ${\bm q}_1={\bm q}+2\pi {\bm e}^1$. All values of $\hat{h}$ are normalized by $\bar{u}$. The set $Q_{\bm q}$ is chosen as ${\bm q}/q=(\sqrt{3}/2,1/2)$ with $q=10^6$ ${\rm m}^{-1}$. (a) $\hat{h}$ as a function of $t$ for ${\bm q}$, ${\bm q}_1$ and ${\bm q}_2={\bm q}+2\pi {\bm e}^2$.  (b) Schematic pictures of the process.}
\label{evolforce} 
\end{figure}

To study this, we choose again $q=10^6\,\mathrm{m}^{-1}$, oriented as before, and we apply a constant force $u_h({\bm q}_1)=\bar{u}$ only to $\hat{h}({\bm q}_1)$ with ${\bm q}_1={\bm q}+2\pi {\bm e}^1\in Q_{\bm q}$ (Fig.~\ref{evolforce}). The corresponding wavelengths of $q$ and ${\bm q}_1$ are then $2\pi/q\approx 6.3\, \mu{\rm m}$ and $2\pi /|{\bm q}_1| \approx 0.87 a$, respectively. We investigate the response of $\hat h$ at the large scale $\bm q$ but also at another small scale, ${\bm q}_2={\bm q}+2\pi {\bm e}^2\in Q_{\bm q}$. With the initial condition $\hat{h}=\rho=0$, all the modes will be proportional to $\bar{u}$. For $t\lesssim1/\gamma_\pm^{(0)}(q_1)$, the small scale deformation $\hat h({\bm q}_1)$ is excited and reaches the stationary value $\hat{h}_{\rm st}({\bm q}_1)\simeq \bar{u}/(\kappa d^2 q_1^4)$ minimizing $F-(2\pi)^{-2}\bar{u}\hat h({\bm q}_1)$.
Then, for $1/\gamma_\pm^{(0)}(q_1) \lesssim t\lesssim \gamma_\pm^{-1}$, the large-scale mode $\hat{h}({\bm q})$ gets excited by $\hat{h}({\bm q}_1)$ via the cytoskeleton deformation, and then for $t\gtrsim\gamma_\pm^{-1}$, $\hat{h}({\bm q})$ reaches the stationary value
\begin{align}
\hat{h}_{\rm st}({\bm q})\simeq -\frac{\Delta\sigma_{\rm eff}+\Delta\kappa_{\rm eff}q^2}{\sigma_{\rm eff}+\kappa_{\rm eff}q^2}\frac{\bar{u}}{\kappa d^2q_1^4},
\end{align}
where $\Delta\sigma_{\rm eff}=\sigma_{\rm eff}-\sigma$ and $\Delta\kappa_{\rm eff}=\kappa_{\rm eff}-\kappa$. With our choice of parameters, $(\Delta\sigma_{\rm eff}+\Delta\kappa_{\rm eff}q^2)/(\sigma_{\rm eff}+\kappa_{\rm eff}q^2) \simeq 1$, and thus $\hat{h}_{\rm st}({\bm q})\simeq -\hat{h}_{\rm st}({\bm q}_1)$, consistent with Fig.~\ref{evolforce}a. We find that the other small-scale modes, such as $\hat h(\bm q_2)$, are also excited, but not significantly (Fig.~\ref{evolforce}a).

When a force distribution is applied to multiple small-scale modes, the magnitude of the excited large-scale mode $\hat{h}({\bm q})$ can become much larger than in the case examined above. To show this, let us consider at each lattice site of the cytoskeleton active forces
inducing some local curvature. Such forces can formally be derived by adding a ``fictitious" potential $U=-\sum_\ell w_\ell\int {\rm d}^2 x \,  \delta({\bm x}-{\bm R}_\ell)\nabla^2 h$ to the free energy. This yields $u({\bm q}')=q'^2 w_{{\bm q}}$, $\forall {\bm q}'\in Q_{\bm q}$ with $w_{\bm q}=\sum_\ell w_\ell e^{-i{\bm q}\cdot {\bm R}_\ell}$. Note that some force is applied also to the large-scale mode ${\bm q}$. By linearity, the effect of the small-scale modes $\bm q'$ on the large-scale mode $\bm q$, denoted by $\delta\hat h(\bm q)$, will be enhanced by a factor $r=\sum'_{m_1,m_2} q_1^2/|2\pi m_\alpha{\bm e}^\alpha|^2$ with respect to the case shown in Fig.~\ref{evolforce} where only the mode $\bm q_1$ was excited. The sum is taken up to the high wavevector cutoff while excluding $m_1=m_2=0$. With the parameters given above, we find $r\approx20$. Assuming that the microscopic forces can produce a deformation of amplitude comparable to the mesh size, i.e., $h_1\simeq50\,\mathrm{nm}$ \cite{BetzPNAS, ParkPNAS}, we expect the large scale response $\delta h$ to be about $r h_1$, which is a sizeable deformation of the order of $1\,\mu\mathrm{m}$. Note that this scale transfer of membrane deformation requires applying the small scale forces for at least about $10\,\mathrm{ms}$ (Fig.~\ref{evolforce}).

\section{Discussion and Summary\label{summary}}
 In this paper, since we assume only vertical motion of the quasi-planer RBC patch, and equilibrium dynamics, we neglect the dissipation due to tangential motion of the cytoskeleton as well as the cytoskeleton activity that was studied in Ref.~\cite{Turlier}. Nevertheless, in Appendix \ref{fric_cyto}, we have considered the friction between the tangential monolayer flow and the anchored proteins, yielding an extra contribution to the lateral force balance equation. However, in our detailed calculation, it is shown to be negligible. As for the viscous drag of the spectrin filaments due to the surrounding fluid, it was also shown to be negligible~\cite{Turlier}. 
For simplicity, we have assumed a quasi-planer membrane, i.e., small deformations about a flat reference shape. 
However, real RBCs are intrinsically curved objects that fluctuate about a curved reference shape \cite{Turlier,EvansPNAS}.
For such cases, not only $\rho=(\rho^+-\rho^-)/2$ but also $\bar\rho=(\rho^++\rho^-)/2$ is coupled to the membrane deformation $h$ \cite{Sachin}. Furthermore, for a curved membrane, the tangential deformation of the cytoskeleton is also coupled to $h$ \cite{Turlier}. For the relaxation of $\bar\rho$, we can show that the inter-monolayer friction is not a dissipation source, while the friction between the anchored proteins and the monolayers is one of the major dissipation sources for large scales satisfying $2\eta q+(\mu+\zeta)q^2\ll \lambda/a^2$. However, in the relaxation of $\bar\rho$ for a bilayer without the cytoskeleton, the inertia effect of the surrounding fluid can not be neglected \cite{Seifert}, so that a more careful study is necessary in the future.

In summary, we have studied the dynamics of RBC membranes modelled as bilayers coupled to a pre-stressed discrete elastic network, and subject to viscous dissipation in the solvent, in each monolayer and between the monolayers. Given the mesh size of the cytoskeleton ($\approx\!100\,\mathrm{nm}$), it is important from the biological point of view to address the dynamics at scales both larger and smaller than the cytoskeleton. Because the latter breaks lateral translational symmetry, each mode is coupled to all the modes that are congruent modulo a wavevector of the cytoskeleton's reciprocal lattice. We have characterized how the small modes renormalize the relaxation rates of the large modes. We have found that, because of the large renormalized tension $\sigma_{\rm eff}$, the shape relaxation dynamics on the large-scales is dominated by the inter-monolayer friction that has regularly been neglected in the previous theories on RBC dynamics \cite{GovPRL, Turlier, EvansPNAS}.
It has been also shown that applying forces on the small scale modes for a sufficiently long time can excite large scale deformations. 

To the best of our knowledge, however, the correlations between different Fourier modes, $\langle h({\bm q}, t) h({\bm q}', t')\rangle$, has not been measured in previous experiments. It is informative to measure this quantity in order to know the precise dynamical processes where modes in different scales are coupled due to the cytoskeleton, and also to understand the behavior of active forces in fresh RBCs.

\begin{acknowledgements}
R.O.~and S.K.~acknowledge support from the Grant-in-Aid for Scientific Research on Innovative Areas “Fluctuation and Structure” (Grant No. 25103010) from the Ministry of Education, Culture, Sports, Science, and Technology of Japan, and from the Grant-in-Aid for Scientific Research (C) (Grant No. 15K05250) from the Japan Society for the Promotion of Science (JSPS).
\end{acknowledgements}

\appendix
\section{Coupling between different Fourier modes \label{couple_mode}}
Notice that the Fourier modes between different wavevectors are coupled with each other because the cytoskeleton network breaks the continuous translational symmetry. More precisely, while ${\cal F}_{\bm q}[\delta F/\delta \rho]=2k[\rho({\bm q}) - dq^2h({\bm q})]$ in Eq.~(\ref{rhodot}) does not couple different Fourier modes, ${\cal F}_{\bm q}[\delta F/\delta h]$ in Eq.~(\ref{hdot}) couples the Fourier modes in the set $Q_{\bm q}=\{ {\bm q}+2\pi m_\alpha {\bm e}^\alpha | m_\alpha  \in \mathbb{Z} \}$. To see this, we rewrite Eq.~(\ref{F_lattice}) as
\begin{align}
F_{\rm c}=\frac{1}{2} \int \d ^2x \frac{\nu}{2} \sum_\ell\sum_{\bm n}\delta({\bm x}-{\bm R}_\ell)[h({\bm x})-h({\bm x}+{\bm n})]^2.
\end{align}
Then we can calculate the Fourier-transformed functional derivative as 
\begin{align}
{\cal F}_{\bm q}\Big[\frac{\delta F}{\delta h}\Big]=&\hspace{1mm}(\sigma q^2+\tilde\kappa q^4) h({\bm q}) -2kdq^2 \rho({\bm q}) \nonumber \\
&+\nu \sum_\ell\sum_{\bm n} e^{-i{\bm q}\cdot{\bm R}_\ell} [ h({\bm R}_\ell)-h({\bm R}_\ell+{\bm n})]. \label{dFdh1}
\end{align}
Now we use the identity
\begin{align}
\sum_\ell e^{-i{\bm q}\cdot {\bm R}_\ell}=\frac{(2\pi)^2}{\sqrt{g}} \sum_{{\bm q}'\in Q_{\bm q}} \delta({\bm q}'), \label{deltaLattice}
\end{align}
where $g=|{\bm e}_\alpha\cdot{\bm e}_\beta|$ is the determinant of the metric tensor in the primitive-vector-frame $\{ {\bm e}_\alpha \}$, and is given by $g=3a^4/4$. Using this identity, we can rewrite Eq.~(\ref{dFdh1}) as
\begin{align}
{\cal F}_{\bm q}\Big[\frac{\delta F}{\delta h}\Big]=&\hspace{1mm}(\sigma q^2+\tilde\kappa q^4) h({\bm q}) -2kdq^2 \rho({\bm q})\nonumber \\
& +\frac{\nu}{\sqrt{g}}\sum_{{\bm q}'\in Q_{\bm q}}h({\bm q}') K_{{\bm q}'}, \label{dFdh2}
\end{align}
where $K_{\bm q}=\sum_{\bm n} (1-e^{i{\bm q}\cdot {\bm n}})$. We can clearly see from Eq.~(\ref{dFdh2}) that in Eq.~(\ref{EQ1}) the Fourier modes in the common set $Q_{\bm q}$ are coupled to each other. Note that without the cytoskeleton $\nu=0$ (and without the force terms $u_h$ and $u_\rho$), Eqs.~(\ref{EQ1}) and (\ref{EQ2}) reduce to the equations studied by Seifert and Langer \cite{Seifert}.

\section{Elimination of the velocity fields\label{elimination_velo}}
In accordance with the in-plane Fourier transform with a wavevector ${\bm q}$, we introduce ${\bm q}_\perp={\bm e}_z \times {\bm q}$, where ${\bm e}_z$ is the unit vector pointing towards the $z$-direction.  Then unit vectors $\hat {\bm q}$ and $\hat {\bm q}_\perp$ are defined as $\hat {\bm q}={\bm q}/q$ and $\hat {\bm q}_\perp={\bm q}_\perp/q={\bm e}_z\times \hat {\bm q}$, respectively. The logitudinal and transverse components of the Fourier transform of ${\bm V}$ and ${\bm v}^\pm$ are defined by
$V_\parallel({\bm q},z)=\hat {\bm q} \cdot {\bm V}({\bm q},z)$, 
$V_\perp({\bm q},z)=\hat {\bm q}_\perp \cdot {\bm V}({\bm q},z)$, 
$v^\pm_\parallel({\bm q})=\hat {\bm q} \cdot {\bm v}^\pm({\bm q})$, 
and $v^\pm_\perp({\bm q})=\hat {\bm q}_\perp \cdot {\bm v}^\pm({\bm q})$.
The Fourier transform of Eq.~(\ref{Stokes}) is written as
\begin{align}
&\eta (\p_z^2-q^2)V_z-\p_z p=0,\\
&\eta (\p_z^2-q^2)V_\parallel-iq p=0,\\
&\eta (\p_z^2-q^2)V_\perp=0, \\
&iqV_\parallel+\p_zV_z=0.
\end{align}
These equations are solved to obtain
\begin{align}
&p= p^\pm e^{\mp qz}, \label{sol1}\\
&V_\perp= v^\pm_\perp e^{\mp qz}, \label{sol2}\\
&V_z=\Big[ \dot{h}+\frac{p^\pm}{2\eta}z\Big]e^{\mp qz}=[ \dot{h} \pm (\dot{h}\mp v^\pm_\parallel )qz] e^{\mp qz}, \label{sol3}\\
&V_\parallel=-\frac{1}{iq}\Big[ \frac{p^\pm}{2\eta}\mp q\Big(\dot{h}+\frac{p^\pm}{2\eta}z\Big) \Big] e^{\mp qz} \nonumber \\
&\hspace{3.5mm}= [v^\pm_\parallel-(i\dot{h}\pm v^\pm_\parallel)qz] e^{\mp qz}, \label{sol4}
\end{align}
where the upper and the lower signs indicate the solutions for $z>0$ and for $z<0$, respectively. 


Substituting Eqs.~(\ref{sol1}) and (\ref{sol3}) into the Fourier transform of Eq.~(\ref{vertical}), we obtain
\begin{align}
4\eta q\frac{\p h({\bm q})}{\p t}=- {\cal F}_{\bm q}\Big[\frac{\delta F}{\delta h}\Big], \label{hdot}
\end{align}
which yields Eq.~(\ref{EQ1}) (without the force term $u_h$).
Next, we use Eq.~(\ref{sol4}) to eliminate $V_\parallel$ and $V_z$ from the longitudinal component of Eq.~(\ref{lateral}), and obtain
\begin{align}
0=&\hspace{1mm}[2\eta q+(\mu+\zeta)q^2] v_\parallel^\pm({\bm q})+iq {\cal F}_{\bm q} \Big[\frac{\delta F}{\delta \rho^\pm}\Big] \nonumber \\
 & \pm b[v_\parallel^+({\bm q})-v_\parallel^-({\bm q})]. \label{vlong}
\end{align}
The Fourier transform of Eq.~(\ref{continuity}) relates $\p \rho/\p t=(\p/\p t) (\rho^+-\rho^-)/2$ with the longitudinal velocity $v_\parallel$ as $\p\rho/\p t=-iq(v_\parallel^+-v_\parallel^-)/2$. Then Eq.~(\ref{vlong}) yields
\begin{align}
0=-\frac{2}{iq}c(q) \frac{\p\rho ({\bm q})}{\p t}+iq {\cal F}_{\bm q} \Big[\frac{\delta F}{\delta \rho^+}-\frac{\delta F}{\delta \rho^-}\Big], \label{rhodot}
\end{align}
where $c(q)=2b+2\eta q+(\mu+\zeta)q^2$. Using the identity $\delta (\cdots) /\delta \rho=\delta (\cdots)/\delta\rho^+-\delta (\cdots)/\delta\rho^-$, we obtain Eq.~(\ref{EQ2}) (without the force term $u_\rho$).

\section{Operator representation and perturbation expansion \label{perturbation}}
We seek the relaxation rates and their associated eigen modes of Eqs.~(\ref{EQ1}) and (\ref{EQ2}) as a power series of $\nu$, since for $\nu=0$ they can be obtained analytically. It is convenient to introduce for each $Q_{\bm q}$ a Hilbert space $S_{\bm q}$ spanned by an orthonormal set $\big\{ |h;{\bm q}'\rangle, \ |\rho;{\bm q}'\rangle \ |\  {\bm q}'\in Q_{\bm q}\big\}$, and its dual space $S_{\bm q}^*$ spanned by an orthonormal set
$\big\{ \langle h;{\bm q}'|, \ \langle\rho;{\bm q}'| \ |\  {\bm q}'\in Q_{\bm q}\big\}$. 
We define the following state vector in $S_{\bm q}$,
\begin{align}
|\Psi(t) \rangle _{\bm q} \equiv \sum_{{\bm q}'\in Q_{\bm q}} [\hat{h}({\bm q}',t)|h;{\bm q}'\rangle+ \rho({\bm q}',t) |\rho;{\bm q}'\rangle  ],
\end{align}
where $\hat{h}=h/d$. Then Eqs.~(\ref{hdot}) and (\ref{rhodot}) (or Eqs.~(\ref{EQ1}) and (\ref{EQ2}) without the force terms $u_h$ and $u_\rho$) are written as
\begin{align}
\frac{\p }{\p t}|\Psi(t)\rangle_{\bm q}= -(\Gamma_{\rm SL}+\nu\Gamma_h)|\Psi(t)\rangle_{\bm q}, \label{main}
\end{align}
where the linear operators $\Gamma_{\rm SL}$ and $\Gamma_h$ are defined by
\begin{align}
\Gamma_{\rm SL}({\bm q})=&\sum_{{\bm q}'\in Q_{\bm q}} \Big[ \frac{kq'^2}{c(q')}  |\rho;{\bm q}'\rangle \langle \rho;{\bm q}'| -\frac{kd^2q'^4}{c(q')}  |\rho;{\bm q}'\rangle \langle h;{\bm q}'| \nonumber \\
& +\frac{\sigma q'+\tilde \kappa q'^3}{4\eta} |h;{\bm q}'\rangle \langle h;{\bm q}'| -\frac{kq'}{2\eta }  |h;{\bm q}'\rangle \langle \rho;{\bm q}'| \Big] ,\label{Lsl}\\
\Gamma_h({\bm q})=& \frac{1}{\sqrt{g}}\sum_{{\bm q}'\in Q_{\bm q}}\sum_{{\bm q}''\in Q_{\bm q}} \frac{K_{{\bm q}'}}{4\eta q''} |h;{\bm q}''\rangle\langle h;{\bm q}'|. \label{Lh} 
\end{align}
Notice that, in the absence of the cytoskeleton ($\nu=0$), Eq.~(\ref{main}) reduces to $0=|\dot \Psi \rangle_{\bm q}+\Gamma_{\rm SL}|\Psi \rangle_{\bm q} $ which was discussed by Seifert and Langer  for vanishing tension $\sigma=0$ \cite{Seifert}.

Let us consider the eigenvalue problem
\begin{align}
(\Gamma_{\rm SL}+\nu\Gamma_h)|e\rangle = \gamma |e \rangle,
\end{align}
where $\gamma$ and $|e\rangle$ are the eigenvalue and the eigenvector of $\Gamma_{\rm SL}+\nu\Gamma_h$, respectively. We expand $\gamma$ and $|e\rangle$ in powers of $\nu$ as $\gamma=\gamma^{(0)}+\nu\gamma^{(1)}+\cdots$ and $|e\rangle=|e^{(0)}\rangle+\nu|e^{(1)}\rangle+\cdots$. The 0th and the 1st order equations read
\begin{align}
&\Gamma_{\rm SL}|e^{(0)}\rangle=\gamma^{(0)}|e^{(0)}\rangle , \label{0th}\\
&(\Gamma_h-\gamma^{(1)})|e^{(0)}\rangle=(\gamma^{(0)}-\Gamma_{\rm SL})|e^{(1)}\rangle. \label{1st}
\end{align}
For the following perturbation calculation, it is convenient to introduce the 0th and 1st order left eigenvectors $\langle e^{(0)\dagger}|$ and $\langle e^{(1)\dagger}|$, respectively. These satisfy
\begin{align}
&\langle e^{(0)\dagger}|\Gamma_{\rm SL}=\langle e^{(0)\dagger} | \gamma^{(0)},\label{0thleft}\\
&\langle e^{(0)\dagger} |(\Gamma_h-\gamma^{(1)})=\langle e^{(1)\dagger}|(\gamma^{(0)}-\Gamma_{\rm SL}) \label{1stleft}. 
\end{align}
Here $\gamma^{(0)}$ and $\gamma^{(1)}$ are common to Eqs.~(\ref{0th}) and (\ref{1st}), respectively. Note that $\langle e^{(0)\dagger} |$ (resp.~$\langle e^{(1)\dagger} |$) is {\it not} the Hermitian conjugate of $|e^{(0)}\rangle$ (resp.~$|e^{(1)}\rangle$), because neither $\Gamma_{\rm SL}$ nor $\Gamma_h$ is a Hermitian operator.

\subsection{0th order}
Since $\Gamma_{\rm SL}$ does not couple the Fourier modes of different wavevectors, we readily obtain the 0th order eigenvalues,
\begin{align}
\gamma^{(0)}_\pm=&\hspace{1mm}\frac{q}{8\eta c(q)} \Big[(\sigma+\tilde\kappa q^2)c(q) \nonumber \\
&+4\eta kq\pm \sqrt{g(q)^2+32\eta k^2d^2 q^3 c(q)}\Big]
\end{align}
with $g(c)=(\sigma+\tilde\kappa q^2)c(q)-4\eta kq$. Their associated right eigenvectors are given by
\begin{align}
|e_\pm^{(0)};{\bm q}\rangle=|h;{\bm q}\rangle+e_\pm(q)|\rho;{\bm q}\rangle,
\end{align}
where $e_\pm(q)=[ g(q)\mp\sqrt{g(q)^2+32\eta k^2 d^2 q^3 c(q)} ]/[4kc(q)]$. The corresponding left eigenvectors are
\begin{align}
\langle e_\pm^{(0)\dagger};{\bm q}|=\frac{1}{1+e_\pm^\dagger(q)e_\pm(q)} \left[ \langle h;{\bm q}|+e_\pm^\dagger(q) \langle \rho;{\bm q}| \right]
\end{align}
with $e_\pm^\dagger(q)=c(q)e_\pm(q)/(2\eta d^2q^3)$. In the above, the left eigenvector is normalized such that $\langle e_\pm^{(0)\dagger};{\bm q}|e_\pm^{(0)};{\bm q}\rangle=1$. Since a contraction of left and right eigenvectors associated with different eigenvalues vanishes, we obtain
\begin{align}
\langle e_\epsilon^{(0)\dagger};{\bm q}'|e_{\epsilon'}^{(0)};{\bm q}''\rangle=\delta_{\epsilon\epsilon'}\delta_{{\bm q}'{\bm q}''}. \label{orth0}
\end{align}
This yields $e_+^\dagger(q)e_-(q)=e_-^\dagger(q)e_+(q)=-1$, which can also be confirmed directly from the definition of $e_\pm$ and $e^\dagger_\pm$.

In Table \ref{TabSL}, we present approximate expressions of $\gamma_\pm^{(0)}$ and $e_\pm$ for different length scales classified by the characteristic wavenumbers $b\sigma/(2k\eta)$ and $q_{\rm c}=2\eta k/(b\tilde\kappa)$, where a small bare tension $\sigma\ll \sigma_{\rm c}$ is assumed \cite{JBNLM}. Typical parameter values quoted in the main text yield $\sigma_{\rm c}\equiv (2\eta k)^2/(\tilde\kappa b^2) \approx 3\times10^{-6}\,\mathrm{J/m}^2$. The behavior in the two regimes (i) and (ii) in Table \ref{TabSL} is essentially the same as those in Ref.~\cite{Seifert} for $\sigma=0$. 
In the presence of tension, there appears another regime $q\ll b\sigma/(2k\eta)$ where the dynamics is again dominated by the inter-monolayer friction \cite{JBNLM}. For $\sigma=10^{-11}{\rm N/m}$ chosen in the main text, the characteristic wavenumber corresponds to the length $4\pi k\eta/(b\sigma) \approx 0.44\, {\rm m}$, which is too large to be measured in experiments. However, with $\sigma=4 \times 10^{-7}{\rm N/m}$, for instance, we have $4\pi k\eta/(b\sigma) \approx 10 \mu{\rm m}$, which is relevant for giant unilamellar vesicles. For large tension $\sigma\gtrsim \sigma_{\rm c}$, on the other hand, the dynamics is dominated by the inter-monolayer friction in all length scales \cite{JBNLM}. 
\begin{table*}[tbh]
\caption{
Approximate expressions of 0th order eigenvalues and eigenvectors for sufficiently small bare tension, $\sigma\ll \sigma_c$.
\label{TabSL}}
\begin{ruledtabular}
\begin{tabular}[t]{l c c c c}
&$\gamma^{(0)}_+(q)$  \quad & $\gamma^{(0)}_-(q)$\ \ \quad  & $e_+(q)$\   \quad & $e_-(q)$\\
\hline
(i) $b\sigma/(2k\eta)\ll q\ll q_{\rm c} $ &$kq^2/(2b)$ & $(\sigma q+\kappa q^3)/(4\eta)$ & $-\eta q/b$ &$(qd)^2$ \\
(ii) $q\gg q_{\rm c}$ & $\tilde\kappa q^3/(4\eta)$  & $k\kappa q^2/(2b\tilde\kappa)$ & $-2k\eta d^2 q/(b\tilde\kappa)$ & $\tilde\kappa q^2/(2k)$ 
\end{tabular}
\end{ruledtabular}
\end{table*}

\subsection{1st order}
We assume for simplicity that the 0th order eigenvalue $\gamma_\pm^{(0)}(q)$ is not degenerated in the Hilbert space $S_{\bm q}$. This assumption is always valid for scales much larger than the lattice spacing of the cytoskeleton, $q\ll a^{-1}$. Equation (\ref{1st}) yields the 1st order correction $\nu\gamma_\pm^{(1)}(q)$ to the 0th order eigenvalue $\gamma_\pm^{(0)}(q)$ as
\begin{align}
\gamma_\pm^{(1)}(q)= \langle e_\pm^{(0)\dagger};{\bm q}| \Gamma_h |e_\pm^{(0)};{\bm q}\rangle =\frac{ K_{\bm q}}{4\sqrt{g}\eta q[ 1+ e_\pm^\dagger(q)e_\pm(q)]}. \label{eigenvalue_1st}
\end{align}
To obtain the shifted eigenvectors, we expand the 1st order eigenvectors in terms of the 0th order eigenvectors,
\begin{align}
 |e_\pm^{(1)};{\bm q}\rangle= \sum_{{\bm q}'\in Q_{\bm q}} \sum_{\epsilon=+,-}s^\epsilon_\pm ({\bm q}';{\bm q}) |e_\epsilon ^{(0)};{\bm q}'\rangle .
\end{align}
Here we set $s^\pm_\pm ({\bm q};{\bm q})=0$ (as for the perturbation theory in quantum mechanics).
Operation of $\langle e_\epsilon^{(0)\dagger}; {\bm q}'|$ to the both sides of Eq.~(\ref{1st}) yields
\begin{align}
s_\pm^{\epsilon}({\bm q}';{\bm q})=&\frac{\langle e_\epsilon^{(0)\dagger}; {\bm q}'| \Gamma_h|e_\pm^{(0)}; {\bm q}\rangle}{\gamma_\pm^{(0)}(q)- \gamma_\epsilon^{(0)}(q') } \nonumber \\
=&\frac{K_{\bm q} }{4 \sqrt{g}\eta q' [\gamma_\pm^{(0)}(q)- \gamma_\epsilon^{(0)}(q')][1+e_\epsilon^\dagger(q') e_\epsilon(q')]}. \label{vec_1st}
\end{align}
Similarly, we can calculate the 1st order left eigenvector $\langle e_\pm^{(1)\dagger};{\bm q}|$ from Eq.~(\ref{1stleft}). One can also show that Eq.~(\ref{orth0}) is generalized to the 1st order in $\nu$ as
\begin{align}
\left(\langle e_\epsilon^{(0)\dagger};{\bm q}'|+\nu \langle e_\epsilon^{(1)\dagger};{\bm q}'|\right) \left(|e_{\epsilon'}^{(0)};{\bm q}''\rangle+\nu |e_{\epsilon'}^{(1)};{\bm q}''\rangle\right) \nonumber \\
=\delta_{\epsilon\epsilon'}\delta_{{\bm q}'{\bm q}''}+O(\nu^2). \label{orth1}
\end{align}

\subsubsection*{Case (i) $q\ll 2\eta k /(b\tilde\kappa)$}
Let us suppose a wavevector ${\bm q}_{\rm L}$ satisfies $q_{\rm L}=|{\bm q}_{\rm L}| \ll 2\eta k /(b\tilde\kappa)$, i.e., case (i) in Table \ref{TabSL}. We then study how the cytoskeleton alters the rates $\gamma_\pm^{(0)}(q_{\rm L})$.   We notice that the 0th order eigenvalue $\gamma_\pm^{(0)}(q_{\rm L})$ is not degenerated in $S_{{\bm q}_{\rm L}}$, because, for such a small wavevector, there is no other wavevector ${\bm q}'\in Q_{{\bm q}_{\rm L}}$ satisfying $|{\bm q}'|=q_{\rm L}$. Using Eq.~(\ref{eigenvalue_1st}) and approximate expressions in Table \ref{TabSL}, we find
\begin{align}
\gamma^{(1)}_+({\bm q}_{\rm L})\simeq \frac{\nu bd^2 K_{{\bm q}_{\rm L}}}{4\sqrt{g}\eta^2},\quad \gamma^{(1)}_-({\bm q}_{\rm L})\simeq \frac{\nu K_{{\bm q}_{\rm L}}}{4\sqrt{g}\eta q_{\rm L}}  ,
 \label{approx1}
\end{align}
when $q_{\rm L}\ll 2\eta k /(b\tilde\kappa)$.
In general, $K({\bm q}_{\rm L})$ depends on the direction of the wavevector $\hat {\bm q}_{\rm L}$. However, it is approximated as 
\begin{align}
K_{{\bm q}_{\rm L}}\simeq \frac{3 (q_{\rm L} a)^2}{2}-\frac{3(q_{\rm L}a)^4}{32}, \label{Kapprox}
\end{align}
for  $q_{\rm L}\ll a^{-1}$,
which is isotropic. We can always use this approximation in Eq.~(\ref{approx1}) which is valid for $q_{\rm L}\ll 2\eta k /(b\tilde\kappa)$. This is because $2\eta k /(b\tilde\kappa)=4.37\times 10^6$ ${\mathrm m}^{-1}$ for typical parameter values chosen in the main text, and it is smaller than the reciprocal of the lattice spacing $a\sim10^{-7}$ ${\mathrm m}$ of the cytoskeleton. Substitution of Eq.~(\ref{Kapprox}) into Eq.~(\ref{approx1}) yields Eq.~(\ref{eigenvalues1}). Note for $\nu=10^{-6}\, {\rm N/m}$ chosen in the main text, $\gamma^{(0)}_-+\nu\gamma^{(1)}_-$ is larger than $\gamma^{(0)}_++\nu\gamma^{(1)}_+$, and therefore by definition, $\gamma_+\simeq \gamma^{(0)}_-+\nu\gamma^{(1)}_-$ and $\gamma_-=\gamma^{(0)}_++\nu\gamma^{(1)}_+$.

Next we discuss the shifted eigenvector $ |e_\pm^{(1)};{\bm q}_{\rm L}\rangle$, $q_{\rm L}\ll 2\eta k /(b\tilde\kappa)$. 
In the present case of $q_{\rm L}\ll 2\eta k /(b\tilde\kappa)$, we can assume $q'\gtrsim 2\pi/a \gg 2\eta k /(b\tilde\kappa)$ for $\forall {\bm q}'\in Q_{{\bm q}_{\rm L}}\setminus\{{\bm q}_{\rm L}\}$ (here ``$\setminus$" indicates set difference). Thus, for ${\bm q}'\neq {\bm q}_{\rm L}$, we can set $\gamma_\pm^{(0)}(q_{\rm L})- \gamma_\epsilon^{(0)}(q')\simeq -\gamma_\epsilon^{(0)}(q')$ in the denominator of Eq.~(\ref{vec_1st}).  Using the expressions in (ii) of Table \ref{TabSL}, we obtain for ${\bm q}'\neq {\bm q}_{\rm L}$,
\begin{align}
\frac{\kappa}{2kd^2}s_\epsilon^-({\bm q}'; {\bm q}_{\rm L}) \simeq s_\epsilon^+({\bm q}'; {\bm q}_{\rm L})\simeq -\frac{ K_{{\bm q}_{\rm L}}}{\sqrt{g}\tilde\kappa q'^4}. \label{vec_approx_large00}
\end{align}
For ${\bm q}'={\bm q}_{\rm L}$, we can assume $\gamma_+^{(0)}(q_{\rm L}) \gg \gamma_-^{(0)}(q_{\rm L})$ in the denominator of Eq.~(\ref{vec_1st}) and obtain
\begin{align}
&\mp [1+e_\pm^\dagger(q_{\rm L})e_\pm(q_{\rm L})]s_\mp^\pm({\bm q}_{\rm L};{\bm q}_{\rm L})\simeq  E({\bm q}_{\rm L}),   \label{vec_approx_large0}
\end{align}
where $E({\bm q}_{\rm L})=   K_{{\bm q}_{\rm L}} /[4\sqrt{g}\eta q_{\rm L}\gamma_+^{(0)}(q_{\rm L})]\simeq b K_{{\bm q}_{\rm L}} /[2\sqrt{g}\eta k q_{\rm L}^3]$.
To make the physical meaning of the shifted eigenvectors clear, we examine the time evolution of the vector
\begin{align}
|\Psi(t)\rangle =\sum_{{\bm q}'\in Q_{{\bm q}_{\rm L}}} \Big[ \hat h({\bm q}',t)|h;{\bm q}'\rangle + \rho({\bm q}',t)|\rho;{\bm q}'\rangle \Big] . \label{state_example}
\end{align}
We may suppose that the modes for $\forall {\bm q}'\in Q_{{\bm q}_{\rm L}}\setminus \{{\bm q}_{\rm L}\}$ decay much faster than the modes for ${\bm q}_{\rm L}$, because in the 0th order $\gamma_\pm^{(0)}(q')$ is much larger than $\gamma_\pm^{(0)}(q_{\rm L})$. Then, after sufficiently large time $t$ satisfying $t\gg \gamma_\pm^{(0)}(q')^{-1}$ for $\forall {\bm q}'\in Q_{{\bm q}_{\rm L}}\setminus \{{\bm q}_{\rm L}\}$, the modes for $\forall {\bm q}'\in Q_{{\bm q}_{\rm L}}\setminus \{{\bm q}_{\rm L}\}$ are in the  quasi-equilibrium state $\hat{h}_{\mathrm{qe}}({\bm q}';\hat{h}({\bm q}_{\rm L}))$ and $\rho_{\mathrm{qe}}({\bm q}';\hat{h}({\bm q}_{\rm L}))$ determined by ${\cal F}_{{\bm q}'} [\delta F/\delta h]={\cal F}_{{\bm q}'} [\delta F/\delta \rho^\pm]=0$ for a given value of $\hat{h}({\bm q}_{\rm L})$.  
Then we may approximate Eq.~(\ref{state_example}) as
\begin{align}
|\Psi(t)\rangle \simeq&\hspace{1mm} \hat h({\bm q}_{\rm L},t)|h;{\bm q}_{\rm L}\rangle + \rho({\bm q}_{\rm L},t)|\rho;{\bm q}_{\rm L}\rangle \nonumber \\
&+\sum_{{\bm q}'\in Q_{{\bm q}_{\rm L}}\setminus\{ {\bm q}_{\rm L}\}} \Big[ \hat h_{\mathrm{qe}}({\bm q}';\hat h({\bm q}_{\rm L},t))|h;{\bm q}'\rangle \nonumber \\
& \hspace{22mm} + \rho_{\mathrm{qe}}({\bm q}';\hat h({\bm q}_{\rm L},t))|\rho;{\bm q}'\rangle \Big] . \label{state_example_app}
\end{align}
This will be justified later (see discussion around Eq.~(\ref{QE})). Using ${\cal F}_{{\bm q}'}[\delta F/\delta \rho]=2k[\rho({\bm q}') - dq'^2h({\bm q}')]$ and Eq.~(\ref{dFdh2}), we obtain Eq.~(\ref{quasi_eq}) at linear order in $\nu$. 
We can then show that the following relation holds up to the first order in $\nu$:
\begin{align}
&|\Psi(t)\rangle \nonumber \\
&\simeq \sum_{\epsilon=+,-}  A_\epsilon(\hat{h}({\bm q}_{\rm L},t), \rho({\bm q}_{\rm L},t)) \Big\{ |e_\epsilon^{(0)};{\bm q}_{\rm L}\rangle +\nu |e_\epsilon^{(1)};{\bm q}_{\rm L}\rangle\Big\} , \label{1st_slowdynamics}
\end{align}
where 
\begin{align}
A_+(\hat{h}({\bm q}_{\rm L}), \rho({\bm q}_{\rm L})) =&\hspace{1mm}\frac{1}{e_--e_+}\Bigg[ \Big\{e_-+\frac{\nu \bar E({\bm q}_{\rm L})}{e_--e_+}\Big\}\hat{h}({\bm q}_{\rm L}) \nonumber \\
& \hspace{3mm}- \Big\{1-\frac{\nu e_-^\dagger \bar E({\bm q}_{\rm L})}{e_--e_+}\Big\} \rho ({\bm q}) \Bigg] ,
\end{align}
and
\begin{align}
A_-(\hat{h}({\bm q}_{\rm L}), \rho({\bm q}_{\rm L})) =&\hspace{1mm}\frac{1}{e_--e_+}\Bigg[ \Big\{1-\frac{\nu  e_+^\dagger\bar E({\bm q}_{\rm L})}{e_--e_+}\Big\}\rho({\bm q}_{\rm L}) \nonumber \\
&\hspace{2mm}- \Big\{e_++\frac{\nu \bar E({\bm q}_{\rm L})}{e_--e_+}\Big\} \hat h ({\bm q}_{\rm L}) \Bigg],
\end{align}
with $\bar E({\bm q}_{\rm L})=2\eta d^2 q_{\rm L}^3 E({\bm q}_{\rm L})/c(q_{\rm L})$ and $e_\pm=e_\pm(q_{\rm L})$.
Therefore, after sufficiently large time $t$, $A_+$ and $A_-$ decay with the rates in Eq.~(\ref{eigenvalues1}). The above discussion indicates that the rates in Eq.~(\ref{eigenvalues1}) and their associated eigenvectors correspond to the relaxation of the (long wavelength) modes of ${\bm q}_{\rm L}$ accompanied by instantaneous relaxation of the other (short wavelength) modes of $\forall {\bm q}'\in Q_{{\bm q}_{\rm L}}\setminus \{{\bm q}_{\rm L}\}$ to the quasi-equilibrium state. Note that even though at initial time we set $\hat{h}({\bm q}')=\rho({\bm q}')=0$ for ${\bm q}'\in Q_{{\bm q}_{\rm L}}\setminus \{{\bm q}_{\rm L}\}$, after sufficiently large time ($t\gg \gamma_\pm^{(0)}(q')^{-1}$) they are {\it excited} by non-zero $\hat{h}({\bm q}_{\rm L})$ to their quasi-equilibrium values in Eq.~(\ref{quasi_eq}). 
\vspace{3mm}

\subsubsection*{Case (ii) $q\gg 2\eta k /(b\tilde\kappa)$}
Next we discuss how the rate $\gamma_\pm^{(0)}(q_{\rm S})$ with  $q_{\rm S}=|{\bm q}_{\rm S}| \gg 2\eta k /(b\tilde\kappa)$ is altered by the cytoskeleton.   In this regime, using the expressions in (ii) of Table \ref{TabSL}, we find
\begin{align}
\gamma_+^{(1)}({\bm q}_{\rm S})\simeq \frac{  K_{{\bm q}_{\rm S}}}{4\sqrt{g}\eta q_{\rm S}},\quad \gamma_-^{(1)}({\bm q}_{\rm S})\simeq \frac{k^2d^2 K_{{\bm q}_{\rm S}}}{\sqrt{g} b \tilde\kappa^2 q_{\rm S}^2}. \label{approx2}
\end{align}
We also examine the magnitude of $\gamma_\pm^{(1)}$ compared with that of $\gamma_\pm^{(0)}$. Using the approximate expressions in (ii) of Table \ref{TabSL}, we obtain
\begin{align}
& \frac{\nu\gamma_+^{(1)}({\bm q}_{\rm S})}{\gamma_+^{(0)}(q_{\rm S})}\simeq \frac{\nu K_{{\bm q}_{\rm S}}}{\sqrt{g}\tilde\kappa q_{\rm S}^4} ,\\
 &  \frac{\nu\gamma_-^{(1)}({\bm q}_{\rm S})}{\gamma_-^{(0)}(q_{\rm S})}\simeq \frac{\nu kd^2 K_{{\bm q}_{\rm S}}}{\sqrt{g} \tilde\kappa \kappa q_{\rm S}^4} \sim \frac{\nu K_{{\bm q}_{\rm S}}}{\sqrt{g}\tilde\kappa q_{\rm S}^4}. \label{ratio}
\end{align}
Therefore the effect of cytoskeleton is negligible if $(aq_{\rm S})^4\gg 2a^2\nu K_{{\bm q}_{\rm S}}/(\sqrt{3}\tilde\kappa) \approx 0.0722 K_{{\bm q}_{\rm S}}$. Since $K_{\bm q}\lesssim 10$, the correction to the rates due to the cytoskeleton is not very large. However, the associated modes do not necessarily decay to zero for $t\gg 1/\gamma_\pm^{(0)}(q_{\rm S})$. In fact, if a large scale Fourier mode exists in $Q_{{\bm q}_{\rm S}}$, i.e., ${\bm q}_{\rm L}\in Q_{{\bm q}_{\rm S}}$ (or equivalently ${\bm q}_{\rm S}\in Q_{{\bm q}_{\rm L}}$), the Fourier modes for ${\bm q}_{\rm S}$ are rather excited to the quasi-equilibrium state by the large scale Fourier mode $\hat{h}({\bm q}_{\rm L})$, as discussed above. We can confirm this explicitly by operating $\langle e_\pm^{(0)\dagger};{\bm q}_{\rm S}|+\nu \langle e_\pm^{(1)\dagger};{\bm q}_{\rm S}|$ to Eq.~(\ref{state_example}). Owing to Eq.~(\ref{orth1}), the resultant quantity decays with the rate $\gamma_\pm^{(0)}(q_{\rm S})+\nu\gamma_\pm^{(1)}({\bm q}_{\rm S})\sim \gamma_\pm^{(0)}(q_{\rm S})$. Hence, for $t\gg\gamma_\pm^{(0)}(q_{\rm S})^{-1}$, we have
\begin{align}
[\langle e_\pm^{(0)\dagger};{\bm q}_{\rm S}|+\nu \langle e_\pm^{(1)\dagger};{\bm q}_{\rm S}|]|\Psi (t)\rangle \simeq 0. \label{QE}
\end{align}
Using the approximations in Table \ref{TabSL}, we can show that Eq.~(\ref{QE}) is equivalent to the quasi-equilibrium condition Eq.~(\ref{quasi_eq}) with ${\bm q}'={\bm q}_{\rm S}$ and ${\bm q}={\bm q}_{\rm L}$.

\section{Effects of friction between the bilayer and the proteins at the cytoskeleton vertices\label{fric_cyto}}
We can take into account the friction between the bilayer and the proteins at the vertices of the cytoskeleton by modifying the lateral force balance Eq.~(\ref{lateral}) to
\begin{align}
0=&-\p_i \left(\frac{\delta F}{\delta\rho^\pm}\right)+\p_j\tau_{ij}^\pm\pm T_{iz}^\pm\mp b(v^+_i-v^-_i) \nonumber \\
&-\lambda \sum_\ell v_i^\pm \delta({\bm x}-{\bm R}_\ell). \label{lateral2}
\end{align}
In the above, the friction coefficient $\lambda$ is assumed to be common for both the upper and lower monolayers. The new term $\lambda \sum_\ell v_i^\pm \delta({\bm x}-{\bm R}_\ell)$ also breaks the translational symmetry, and hence leads to the coupling between different Fourier modes. With this new term, Eq.~(\ref{rhodot}) is modified to
\begin{align}
0=&-\frac{2}{iq}c(q) \frac{\p\rho ({\bm q})}{\p t}+iq {\cal F}_{\bm q} \Big[\frac{\delta F}{\delta \rho^+}-\frac{\delta F}{\delta \rho^-}\Big] \nonumber \\
&+\lambda\hat{\bm q}\cdot{\cal F}_{\bm q}\left[\sum_\ell ({\bm v}^+-{\bm v}^-)\delta({\bm x}-{\bm R}_\ell)\right]. \label{rhodot2}
\end{align}
With the use of the identity Eq.~(\ref{deltaLattice}), the new term is rewritten as
\begin{align}
&{\cal F}_{\bm q}\left[\sum_\ell ({\bm v}^+-{\bm v}^-)\delta({\bm x}-{\bm R}_\ell)\right] \nonumber \\
&=-\frac{2}{i\sqrt{g}} \sum_{{\bm q}'\in Q_{\bm q}} \frac{1}{q'^2} [ {\bm q}'\dot\rho({\bm q}')+{\bm q}'_\perp w_\perp ({\bm q}')], \label{fricFourier}
\end{align}
where $w_\perp=-iq( v_\perp^+-v_\perp^-)/2$. To eliminate $w_\perp$ from Eq.~(\ref{rhodot2}), we need the transverse part of Eq.~(\ref{lateral2}). As in a similar way to derive Eq.~(\ref{rhodot2}), we obtain
\begin{align}
0=&-\frac{2}{iq} c_w(q)w_\perp({\bm q}) \nonumber \\
&+\lambda\hat{\bm q}_\perp\cdot{\cal F}_{\bm q}\left[\sum_\ell ({\bm v}^+-{\bm v}^-)\delta({\bm x}-{\bm R}_\ell)\right] \label{wperp}
\end{align}
with $c_w(q)=2b+\eta q+\mu q^2$. Equations (\ref{hdot}), (\ref{rhodot2}) and (\ref{wperp}) are the complete set of the relaxation equations for $\rho$ and $\hat{h}$. We can see from Eq.~(\ref{fricFourier}) that the Fourier modes for ${\bm q}$ are coupled to the Fourier modes for $\forall {\bm q}'\in Q_{\bm q}$, as in the case without the friction at the vertices of the cytoskeleton. 

The full equations are also represented in terms of operators and vectors in the Hilbert space $S_{\bm q}$, as in the previous section. To perform perturbation calculations, we regard both $\nu$ and $\lambda$ as small parameters. Then, to the first order in $\nu$ and $\lambda$, the governing equation is
\begin{align}
\frac{\p }{\p t}|\Psi(t)\rangle_{\bm q}= -(\Gamma_{\rm SL}+\nu\Gamma_h-\lambda\Gamma_\rho \Gamma_{\rm SL})|\Psi(t)\rangle_{\bm q}, \label{main2}
\end{align}
where the operator $\Gamma_\rho$ is defined as
\begin{align}
\Gamma_\rho({\bm q})=\frac{1}{\sqrt{g}}\sum_{{\bm q}'\in Q_{\bm q}}\sum_{{\bm q}''\in Q_{\bm q}}  \frac{{\bm q}''\cdot {\bm q}'}{c_w(q'') q'^2 }|\rho;{\bm q}''\rangle\langle\rho;{\bm q}'|. \label{Lrho}
\end{align}
The correction to the rate $\gamma_\pm^{(0)}(q)$ due to the friction is then given by
\begin{align}
D_\pm(q)\equiv& -\lambda \langle e_\pm^{(0)\dagger};{\bm q}|\Gamma_\rho\Gamma_{\rm SL}|e_\pm^{(0)};{\bm q}\rangle \nonumber \\
 =&-\frac{\lambda \gamma_\pm^{(0)}(q) e_\pm^\dagger(q)e_\pm(q)}{\sqrt{g}c_w(q)[1+e_\pm^\dagger(q)e_\pm(q)]}.
\end{align}
 To measure the relevance of $D_\pm(q)$, we shall consider the ratio $|D_\pm/\gamma_\pm^{(0)}|$. Using the approximations in (i) $b\sigma/(2k\eta)\ll q\ll 2\eta k/(b\tilde\kappa)$ of Table \ref{TabSL}, we obtain 
 \begin{align}
&\left| \frac{D_+(q)}{\gamma_+^{(0)}(q)}\right|\simeq \frac{\lambda}{2\sqrt{g} b}\sim \frac{\lambda}{ba^2} \\
 &\left| \frac{D_-(q)}{\gamma_-^{(0)}(q)}\right| \simeq \frac{\lambda d^2 q}{2\sqrt{g} \eta} \ll \frac{\lambda kd^2}{\sqrt{g} \tilde\kappa b}\sim \frac{\lambda}{\sqrt{g}b}\sim \frac{\lambda}{ba^2}.
\end{align}
Similarly, with the approximations in (ii) $q\gg 2\eta k/(b\tilde\kappa)$ in Table \ref{TabSL}, the ratio $|D_\pm/\gamma_\pm^{(0)}|$ is comparable or smaller than $\lambda/(ba^2)$. Therefore the friction due to the network is negligible as long as the friction coefficient per area $\lambda/a^2$ is much smaller than the coefficient $b$ for the inter-monolayer friction. We use Saffman--Delbr\"uck theory to estimate the value of $\lambda$ \cite{SD}. In this theory two physical situations are examined; (i) membrane of finite size with surrounding fluid being neglected, and (ii) membrane of infinite size with surrounding fluid being taken into account. Since $\lambda$ is a  ``bare" friction constant between a protein and a monolayer with surrounding fluid being neglected, we may use the result for (i). Then we can set  $\lambda\simeq4\pi\mu/\ln(L/r_0)$, with $L$ the membrane size and $r_0$ the protein size. Using $L/r_0\simeq10^2$, $b\simeq2\times 10^8\,\mathrm{J\,s}/\mathrm{m}^4$, $\mu\simeq 2\times 10^{-9}\,\mathrm{J\,s}/\mathrm{m}^2$ and $a\simeq10^{-7}$\,m, we estimate $\lambda/(a^2b)\simeq 5\times 10^{-3}$. We can thus neglect the effects of $\lambda$. 
Note that the estimation with (ii) also leads to the same conclusion; for (ii), we set $\lambda\simeq 4\pi \mu/\ln[\mu/(\eta r_0)]$, and obtain $\lambda/(a^2b)\simeq 1.8\times 10^{-3}$ when $r_0\simeq 2\times 10^{-9}\, \mathrm{m}$.




%
\end{document}